\documentclass[12pt]{iopart}
\usepackage{psfig,epsfig,epsf}
\begin{document}

\title{Resonant enhancement of the jump rate in a double-well
potential}

\author{ Y Zolotaryuk\dag,\ddag,\footnote[3]{To whom correspondence should be 
addressed (yzolo@bitp.kiev.ua)}, V N Ermakov\dag \
and P L Christiansen\ddag }

\address{
 \dag\ Bogolyubov Institute for Theoretical Physics,
National Academy of Sciences of Ukraine,
vul. Metrologichna 14B,
03143 Kyiv, Ukraine}

\address{
\ddag\ Section of Mathematical Physics, IMM,
Technical University of Denmark, DK-2800 Lyngby, Denmark
}

\begin{abstract} 
We study the overdamped dynamics of a Brownian particle
in the double-well potential under the influence of an
external periodic (AC) force with zero mean.
We obtain a dependence of the jump rate on the frequency of the
external force. 
The dependence shows a maximum at a certain driving frequency.
We explain the phenomenon as a switching
between different time scales of the system: interwell relaxation time
(the mean residence time) and the intrawell relaxation time.
Dependence of the resonant peak on the system parameters, namely
the amplitude of the driving force $A$ and the noise strength 
(temperature) $D$ has been explored. We observe that
the effect is well pronounced when $A/D>1$ and if
$A/D \gg 1$ the enhancement of the jump rate can be of the
order of magnitude with respect to the Kramers rate.

\end{abstract}

 
\submitto{\JPA}

\maketitle

\section{Introduction and the background}
In the last decades a lot of interest  has been devoted toward
the problem of the behaviour of a nonlinear system under the combined influence
of stochastic and time-periodic forces. A number of remarkable 
phenomena such as stochastic resonance (SR) and resonant activation 
 has been discovered and extensively studied 
both experimentally and theoretically\cite{gjhm98rmp,anms-g99ufn,k99ufn}.
So far, a lot of experimental confirmations of this effect has been
reported in different areas of physics, like optics, biophysics or
condensed matter to name a few. 
In our opinion, the frequency dependence of the SR is one
of the most intriguing problem in this area. Initially
defined as a problem of {\it ``bona fide''} 
stochastic resonance\cite{gms95prl}, it has created a vivid
discussion in scientific literature\cite{zmj90pra,gms95prl}. The main
question in this discussion can be formulated as follows.
It is a common knowledge, that the SR phenomenon manifests
itself as a maximum of the signal-to-noise ratio (SNR)
at the temperature, for which the Kramers rate equals the 
double driving frequency. However it is not possible
to display a similar resonant characteristic as a function
of frequency. 

On the other hand, experimental observation of the 
{\it resonant escape} from the potential well  \cite{dmec84prl}
has stimulated a number of theoretical papers \cite{theor,smas01pre}.
It has been shown that an {\it underdamped} particle
driven by both noise and an AC force can perform a
resonant escape from the potential well at a certain resonant
frequency.  However, the narrow resonant peak has been obtained only for
underdamped or moderately damped systems, and only recently 
Pankratov and Salerno in \cite{ps00pla}
have shown that the mean transition time of an overdamped particle over
a periodically oscillating potential well can exhibit a resonant 
dependence on the frequency of this oscillation in the
case of strong (comparing to the depth of the well) driving forces.

The problem we would like to present in this paper
is whether an enhancement of the well-to-well transfer (jump) rate is 
possible in the
{\it overdamped} bistable system for an arbitrarily 
strong (or weak) AC force. 
We believe that the answer to this question is also important
because the unperturbed overdamped system does not have oscillatory dynamics,
and, therefore does not possess oscillatory frequencies.
We are also interested in finding the
frequency dependence of the average jump rate for different
values of noise and driving amplitude.

\section{The model}

\begin{figure}[htb]
\begin{center}
\epsfxsize=3.in
\centerline{\epsfbox{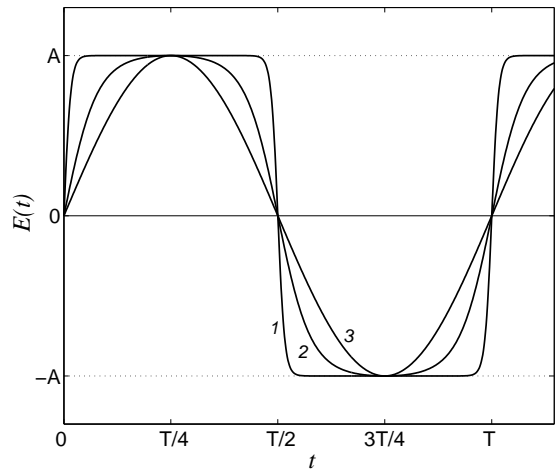}}
\end{center}
\caption{Shape of the driving force $E(t)$ for different values of the
parameter $\mu$: $\mu=10$ (curve 1),  $\mu=2$ (curve 2) and $\mu=0$ 
(curve 3).}
\label{fig0}
\end{figure}

Consider the overdamped particle in a double-well
potential $V(x)$ and under the action
of a periodic external force and a white noise.
The time evolution of the particle position $x(t)$ is 
governed by the Langevin equation 
%--------------------------------------------------------------------- 
\begin{equation}
 \gamma \dot{x} = F(x,t) + \xi(t),\; F(x,t)= -V'(x)+ E(t).
\label{1}
\end{equation}
%---------------------------------------------------------------------
Here the dot stands for differentiation with respect to time, 
$\gamma$ is the dissipation (damping) parameter, 
and the double-well potential is chosen as in the $\phi^4$ model:
$V(x)=(x^2-1)^2/4$. The relaxation time scale is given by 
$t_{relax}=2\gamma/V''(\pm 1) \equiv \gamma$. The driving 
force $E(t)=E(t+T)$ has been taken as
%====-----------------------------------------------------------------
\begin{equation}
E(t)= A \tanh {[\mu \sin (\omega t)]}/\tanh {\mu}.
\end{equation} 
%---------------------------------------------------------------------
The parameter $\mu$ controls the shape of the function $E(t)$ 
(see Fig. \ref{fig0} for details). In the limit
$\mu=0$ we obtain simple sinusoidal drive, $E(t)=A \sin (\omega t)$. In the
opposite limit, $\mu \rightarrow \infty$, the driving force is of the 
stepwise form, e.g. $E(t)=+A$ or $-A$ during each of the halfperiods 
of the driving force. 
%@@@@@@@@@@@@@@@@@@@@@@@@@@@@@@@@@@@@@@@@@@@@@@@@@@@@@@@@@@@@@@@@@@@@@@
%
%-------------------Figure 1-------------------------------------------
%%@@@@@@@@@@@@@@@@@@@@@@@@@@@@@@@@@@@@@@@@@@@@@@@@@@@@@@@@@@@@@@@@@@@@@@
The importance of the shape of the driving force
will be discussed later. The AC force is assumed to be weak, 
e.g. $A \ll 1$, so that the double-well system does not turn itself
into a single-well system during the oscillations of $E(t)$.

The temperature effect is given by a zero-mean stochastic force
 $\langle \xi(t) \rangle =0$ and auto-correlation function
\begin {equation}
 \langle \xi(t)\xi(t')\rangle=2D\gamma\, \delta(t-t'),
\end {equation}
with $D$ being the strength of the noise (dimensionless temperature).
Similarly to Eckman and Thomas\cite{et82jpa} we are going to 
characterize the phenomenon by the average number of jumps (or the jump
rate) from one well to the other one:
%---------------------------------------------------------------------
\begin{equation}
\bar n  = \lim_{\tau \rightarrow \infty} \frac{N_j}{\tau},
\end{equation}
%---------------------------------------------------------------------
where $N_j$ is the number of jumps during the time interval $\tau$.

In the absence of the periodic driving ($A=0$) the average jump
rate is defined by the well-known Kramers formula (see, for
example, \cite{risken}). Note that for our choice of the potential
the height of the barrier is $\Delta V=V(x_{max})-V(x_{min})=1/4$,
and, as a result, the Kramers rate equals
%---------------------------------------------------------------------
\begin{equation}
 r_k=
\frac{1}{\sqrt{2}\pi\gamma }\exp {\left (-\frac{1}{4D}\right )}
\left [1-3D/2+{\cal O}(D^2) \right ].
\label{5}
\end{equation}
%---------------------------------------------------------------------
We introduce the amplification factor $\eta$ as a ratio of the
average number of jumps in the presence of the driving fields
to the jump rate when it is absent: 
$\eta(\omega)={\bar n}(\omega)/r_k$.
Since it is not possible to solve the corresponding Fokker-Planck
equation analytically for all ranges of the driving frequency,
we will tackle our problem numerically.

\section{Resonant nature of the jump rate}

We integrate the
corresponding Langevin equation (\ref{1}) with the Runge-Kutta method and
compute numerically the number of particle jumps from one well
to another one. We assume that the jumping event took place when the
particle had reached the opposite well of the potential.
The numerical experiment has been performed during the long times
$\tau \gg T=2\pi/\omega,1/r_k$, $t_{relax}=\gamma$, so that the time 
averaging will  eliminate any possible dependence on the initial phase.

\subsection{Main results of the numerical simulations}

\begin{figure}[htb]
\begin{center}
%\centerline{\epsfig{file=fig1.eps,width=2.6in,height=2.0in,angle=0}}
%\leavevmode
\epsfxsize=4.in
\centerline{\epsfbox{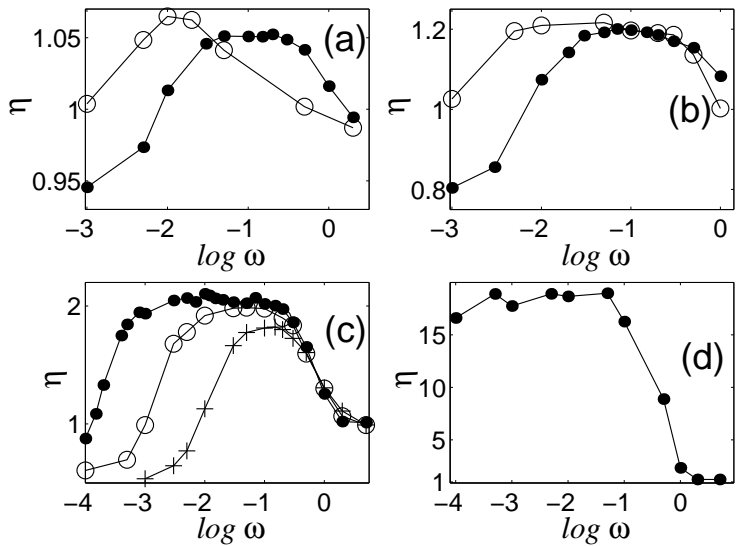}}
\end{center}
\caption{Dependence of amplification of the jump rate $\eta$ 
on the logarithm of the
driving frequency for $\mu=0$, $\gamma=1$, and different ratios $A/D$.
Panel (a): $A/D=1/2$; $D=0.04$($\circ$),$D=0.06$($\bullet$).
Panel (b): $A/D=1$; $D=0.04$($\circ$), $D=0.06$($\bullet$).
Panel (c): $A/D=2$; $D=0.03$($\bullet$), $D=0.04$ ($\circ$),
$D=0.06$ ($+$).
Panel (d)  $A/D=5$, $D=0.02$.}
\label{fig2}
\end{figure}

Numerical simulations clearly show resonant behaviour of the 
amplification factor $\eta(\omega)$.
In Fig. \ref{fig2} we show the dependence  $\eta$ on the 
driving frequency for the different values of $A$  and $D$. 
%@@@@@@@@@@@@@@@@@@@@@@@@@@@@@@@@@@@@@@@@@@@@@@@@@@@@@@@@@@@@@@@@@@@@@@
%
%-------------------Figure 2-------------------------------------------
%%@@@@@@@@@@@@@@@@@@@@@@@@@@@@@@@@@@@@@@@@@@@@@@@@@@@@@@@@@@@@@@@@@@@@@@
Each of the four panels in the figure corresponds
to the certain ratio $A/D$. The resonant nature of the phenomenon
can be seen for all values of $A/D$, however it attains considerable
values if $A/D \ge 1$. For example, in the panel (d) we present the 
case of $A/D=5$ where the
enhancement can reach the values of $10\div 15$. Note that we did not 
continue down to the values $\omega < 10^{-4}$ since
it will require too long averaging times. However, even in the case (d)
maximum of the dependence $\eta=\eta(\omega)$ can be seen.

%@@@@@@@@@@@@@@@@@@@@@@@@@@@@@@@@@@@@@@@@@@@@@@@@@@@@@@@@@@@@@@@@@@@@@@
%
%-------------------Figure 3-------------------------------------------
\begin{figure}[htb]
\begin{center}
\leavevmode
\epsfxsize=4.in
\epsfbox{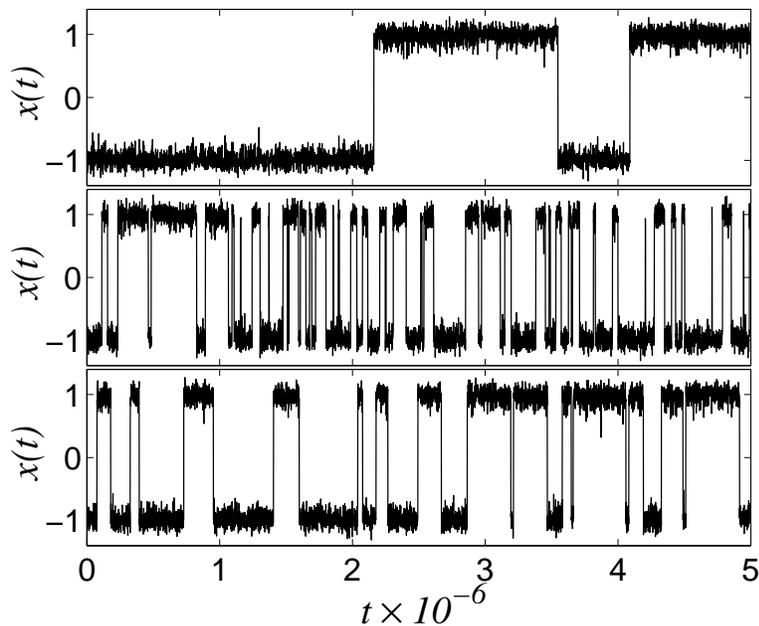}
\end{center}
\caption{Three examples of time evolution of the dynamic variable $x(t)$ 
for $\gamma=1$, $\mu=0$, $D=0.02$ at
$A=0$, $\omega=0$ (upper panel); $A=0.1$, $\omega=0.001$ (middle panel) 
and $\omega=0.5$ (lower panel).}
\label{fig3}
\end{figure}
%%@@@@@@@@@@@@@@@@@@@@@@@@@@@@@@@@@@@@@@@@@@@@@@@@@@@@@@@@@@@@@@@@@@@@@@

Particulars of the dynamics for different values of the driving
frequency are shown in Fig. \ref{fig3}. The upper panel
shows the time evolution of the particle in the case when no 
periodic driving force is present. Here jumping events are
extremely rare and occur at the rate of order of  
$r_k \sim 8 \times 10^{-7}$ [given by the Kramers formula (\ref{5})]. 
The middle panel shows the particle dynamics when the driving 
frequency equals 
$\omega=0.001$, that corresponds to the frequencies close to the
resonance. Gross enhancement of jumping events is clearly seen. The
last panel deals with the case of frequencies much higher than the
resonant frequency ($\omega=0.5$ in this case) where the
intensity of jumping events is again low.

\subsection{Resonance in the jump rate as a switching between different
time scales}

There are three important time scales in our system: the period
of the driving force $T=2\pi/\omega$, the mean residence time in the well, 
inverse to the Kramers rate, $1/r_k$ and the relaxation time of the system, 
$t_{relax}=\gamma\ll 1/r_k$. It is not difficult to see that the dependencies
$\eta=\eta(\omega)$, shown in Fig. \ref{fig2},  consist of three distinct parts:
(i) $\eta(\omega)$ increases in the adiabatic 
limit ($\omega \ll r_k$) from zero frequencies and
continues to grow when $\omega$ exceeds the value of the Kramers rate $r_k$;
(ii) the plateau in the intermediate area $r_k < \omega < 1/\gamma$,
where the change of $\eta$ is slow  and (iii) the high frequency limit, 
$\omega  \gg 1/\gamma$ where $\eta$ decreases with $\omega$ until  $\eta$ 
reaches unity. We shall try to explain the observed resonant enhancement of the
jump rate the switching of the above timescales.

First, we try to explain why the jump rate grows in the adiabatic limit.
Initially we consider the case of sinusoidal drive ($\mu=0$). 
In the adiabatic limit $\bar n \rightarrow r_k$ only 
for very small driving amplitudes ($A \rightarrow 0$). 
 Momentary escape rates from the lower ($r_-$) and upper wells 
($r_+$) have been calculated in \cite{j93pr}:
%======================================================================
\begin{equation}
r_{\pm}(\phi) \simeq r_k \xi_{\pm}(A,\phi) \exp 
\left [ \pm \frac {A \sin \phi}{D} \xi_{\pm}(A,\phi) \right ],~
\xi_{\pm}(A,\phi)=1 \mp \frac{3A}{4}\sin\phi ,
\label{6}
\end{equation}
%======================================================================
and are, in principle, functions of the the phase $\phi=\omega t$. The
momentary residence times $t_{\pm}(\phi)=1/r_{\pm}(\phi)$ obviously also
depend on the phase. It is easy to see
that for the same phase $\phi$ the increase of the average residence 
time in the lower well (${\bar t}_-$)
 does not compensate the decrease on the average residence time in
the upper well (${\bar t}_+$). In other words, 
${\bar t}_- + {\bar t}_+ \neq 2/r_k$, 
and this ``asymmetry'' in average residence times must 
depend on the driving amplitude $A$ in the nonlinear fashion. Using
Eq. (\ref{6}), we show the following:
%======================================================================
\begin{equation}
\frac{2}{t_-(\phi) +t_+(\phi)}=\frac{2r_-(\phi)r_+(\phi)}{r_-(\phi)+
r_+(\phi)}=
r_k\frac{(1-\frac{9A^2\sin^2 {\phi}}{16})\exp(-\frac{3A^2\sin^2\phi}{4D})}
{\cosh \left (\frac{A\sin \phi}{D}\right )
-\frac{3A\sin\phi}{4} \sinh \left (\frac{A\sin \phi}{D}\right )}.
\label{6a}
\end{equation}
%======================================================================
After some simple manipulations one can obtain that for any 
physically reasonable values $A$ and $D$ and for any $\phi$ except 
$\phi = 0, \pm \pi$ the expression multiplied by
$r_k$ in Eq. (\ref{6a}) will be less then $1$. Thus, since
the inequality $ t_-(\phi) +  t_+(\phi)  \neq 2/r_k$ holds, the 
same should hold for the average values: 
${\bar t}_- + {\bar t}_+ \neq 2/r_k$. 
The average number of jumps from well to well can be treated as the 
inverse mean residence time in the 
well, $\bar t=({\bar t}_- + {\bar t}_+)/2\equiv 1/\bar n$. Strictly speaking, 
to compute  the mean residence times in the upper
and lower wells,  ${\bar t}_{\pm}$, one must perform proper averaging
over the phase $\phi$ along the interval $0\le \phi \le \pi$. Notice that
the greatest contribution to the asymmetry in the jump
rates comes from the times when the driving force $E(t)$ attains its
extreme values, $t=nT/4$, $n=1,3,5,...$.
 And, on contrary, at the zeroes of $E(t)$ [$t=nT/2$, $n=0,1,2,...$] 
the potential is symmetric and their contribution to the asymmetry of 
residence times is minimal. Thus, the following estimate of the 
average jump rate (or of the amplification factor $\eta$) is possible:
%======================================================================
\begin{equation}
\frac{2}{r_k[t_-(0) +t_+(0)]}=1>\eta> 
\frac{2}{r_k[t_-(\frac{\pi}{2}) +t_+(\frac{\pi}{2})]}=
\frac{(1-\frac{9A^2}{16})
\exp(-\frac{3A^2}{4D})}{\cosh \left (\frac{A}{D}\right )
-\frac{3A}{4} \sinh \left (\frac{A}{D}\right )}.
\label{7}
\end{equation}
%======================================================================
Thus we can conclude that $\eta(\omega \rightarrow 0)<1$. Also, it is 
straightforward to show that 
$\eta \rightarrow 1$ if $A \rightarrow 0$ and $\eta$ 
decreases with growth of $A$. This is in accordance with numerical
results of  Fig. \ref{fig2}, where in panels (a)-(c) it is clearly
seen that $\eta<1$ when the driving frequency tends to zero.
Also from the numerical results 
of Fig. \ref{fig2} one notices that with the growth of the 
driving frequency the amplification ratio starts to increase.  
In the adiabatic limit (when $T \rightarrow \infty$) 
the asymmetry of the residence times in the upper and lower wells is 
maximal. As long as we increase the frequency (decrease
$T$), less jumping events can take place during one driving period.
As a result, more often the jumping events will take place around
the zeroes of $E(t)$, and, consequently, 
the contribution of the extreme values of $E(t)$ into
 the asymmetry between residence times must
decrease. This must cause the growth of the jump rate. 
However, the most remarkable fact is  that the amplification
ratio $\eta$ does not stop growing when it reaches unity, but 
continues to increase.

Another limit of our problem is the limit of 
high driving frequencies $\omega \gg 1/\gamma$. In this
case, as opposing to the adiabatic case, a lot of oscillations of
the driving force can take place in between the two consecutive 
jumping events. Thus, the system does not ``feel''
the AC force on average, and, since $\langle E(t)\rangle_t=0$, the
influence of the AC drive diminishes.
Finally, in the limit $\omega \rightarrow \infty$ the jump rate
must tend to the Kramers rate. 
This is in accordance with previous studies \cite{j93pr,sdg99prl} 
of the resonant escape from the potential well. 

\begin{figure}[htb]
\begin{center}
\leavevmode
\epsfxsize=4.5in
\epsfbox{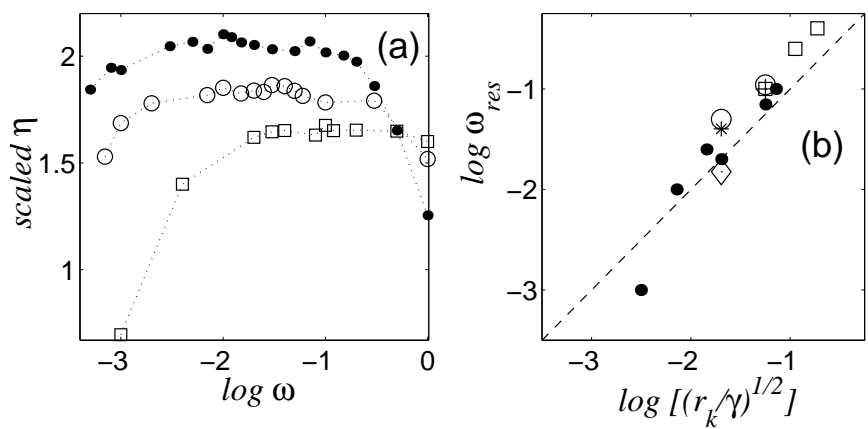}
\end{center}
\caption{Panel (a). Dependence of the scaled amplification factor on 
the driving frequency
at $A=0.06$, $D=0.03$ and $\mu=0$
for different values of damping: $\gamma=1$ ($\bullet$), $\gamma=0.5$($\circ$), 
scaled by 90$\%$,
and $\gamma=0.1$ (squares), scaled by 80$\%$.\\
Panel (b). The resonant frequency $\omega_{res}$ as a function of 
$(r_k/\gamma)^{1/2}$ for $A=0.02$ ($\diamond$), 
$A=0.03$ ($+$), $A=0.04$ ($*$), $A=0.06$ ($\bullet$),
$A=0.08$ ($\circ$) and $A=0.1$ (squares). The dashed line is 
the bisectrix of the straight angle
and is shown as a guide to an eye. }
\label{fig5}
\end{figure}

Thus, we can conclude that
there must exist an optimal driving regime for which the Brownian particle jumps
between the wells with the maximal intensity, and this regime must be somewhere
in between the adiabatic and high-frequency limits, e.g. in the interval
$r_k < \omega < 1/\gamma$.
We have estimated numerically the value of the resonant 
frequency as $\omega_{res} \simeq (r_k/\gamma)^{1/2}$ and
this value is much larger then the Kramers rate. In Fig. \ref{fig5} 
we investigate this
dependence in more detail. In the panel (a) we observe that the maximum
of the dependence $\eta=\eta(\omega)$ shifts to the right when $\gamma$ is
decreased (while all other parameters are fixed). In the panel (b) we 
present the behaviour of the resonant frequency $\omega_{res} $ and  
we see that it does not show any clear dependence on the amplitude.
%@@@@@@@@@@@@@@@@@@@@@@@@@@@@@@@@@@@@@@@@@@@@@@@@@@@@@@@@@@@@@@@@@@@@@@
%
%-------------------Figure 4-------------------------------------------
%%@@@@@@@@@@@@@@@@@@@@@@@@@@@@@@@@@@@@@@@@@@@@@@@@@@@@@@@@@@@@@@@@@@@@@@
Although we do not have a rigorous theory to support such a formula, 
qualitatively it is easy to understand that this behaviour of the resonant
frequency is a result of two competing effects: increasing of the 
temperature tends to activate the jumping process while damping
works against it. 

From this estimate of $\omega_{res}$ we see that the resonant
frequency is linked with the value of the Kramers rate. Doubling the 
temperature may increase $\omega_{res}$ by an order of 
magnitude. Moreover, the steep growth of the curve 
$\eta(\omega)$ depends solely on
the Kramers rate, as it is very clearly demonstrated 
in Fig. \ref{fig2}c for three different values of $D$ and in 
Fig. \ref{fig5}a for three different values of $\gamma$. 
The position of the inflection
point in the curve $\eta(\omega)$, which approximately separates the areas
of the steep growth and the plateau, moves to the right when the temperature $D$ is
increased. This inflection
point is always greater then $r_k$ but is approximately the same order of magnitude 
as $r_k$. On the 
other hand, decrease of the dependence $\eta(\omega)$
is controlled only by the damping parameter. 
It is easy to see that the inflection
point which separates the plateau from the high-frequency area does
not depend on the temperature. However, 
it depends strongly on damping $\gamma$ (see Fig. \ref{fig5}) and matches 
approximately the relaxation frequency $1/t_{relax}=1/\gamma$ 
(see Figs. \ref{fig2}). This can be explained 
in the following way: the most probable dynamics is that the 
particle jumps from the upper well to the lower well. If the time during 
which the particle moves from one well to another one is of the order of
$T/2$, the particle will appear again in the upper well and will
perform a jump with a higher probability again. When $\omega$ increases,
such synchronization with the relaxation frequency will be lost and
if $\omega \gg 1/\gamma$ the particle will have no time to relax in between
the oscillations of the potential. Thus, the jumping events will be more rare,
the jump rate will decrease. As we enter the high-frequency area the system
will react on the external periodic field very weakly.
We can conclude that in order to get the maximal jump rate, the frequency of
the drive must be synchronized with {\it} both the Kramers rate $r_k$ and 
the characteristic relaxation frequency $1/\gamma$.

\subsection{Resonant jump rate as a nonlinear response}

\begin{figure}[htb]
\begin{center}
%\centerline{\epsfig{file=fig3.eps,width=2.6in,height=2.0in,angle=0}}
\leavevmode
\epsfxsize=2.8in
\epsfbox{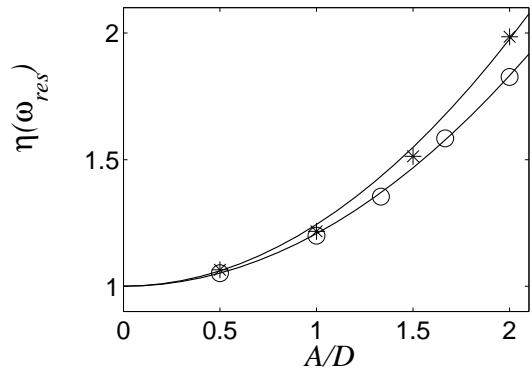}
\end{center}
\caption{Dependence of the amplification ratio at the resonant value
of the driving frequency on the ratio $A/D$ for $\gamma=1$ and 
$\mu=0$: $D=0.04$ ($*$) and 
$D=0.06$ ($\circ$).
Solid lines are the best fits by the parabolic function (see text for 
details).}
\label{fig6}
\end{figure}

An important question is how the observed phenomenon depends
on the amplitude of the driving force, $A$. In this subsection
we show that this dependence has the nature of nonlinear response
to the driving force. In principle, there are many ways to demonstrate
this. As one example, in Fig. \ref{fig6}
we have plotted the dependence of the amplification factor at its
maximal value as a function of the ratio $A/D$ with $D$ being fixed for
each of two curves, shown in the figure. Both curves have been fitted
%@@@@@@@@@@@@@@@@@@@@@@@@@@@@@@@@@@@@@@@@@@@@@@@@@@@@@@@@@@@@@@@@@@@@@@
%
%-------------------Figure 5-------------------------------------------
%@@@@@@@@@@@@@@@@@@@@@@@@@@@@@@@@@@@@@@@@@@@@@@@@@@@@@@@@@@@@@@@@@@@@@@
with the parabolic functions. For the case $D=0.04$  we have obtained 
$1+0.245(A/D)^2$ and for $D=0.06$ we have 
obtained $1+0.21(A/D)^2$. Another example which supports 
the above statement about the nonlinear response is 
the behaviour of the lower boundary of the amplification ratio $\eta$ 
in the adiabatic limit [see Eq. (\ref{7})]. 
The expansion of this formula in the 
powers of $A/D$ yields $1-(17/16)(A/D)^2+O((A/D)^4)$. These results 
unambiguously show that resonant enhancement of the jump rate
can not be described within the framework of the linear
response.

\subsection{Dependence on the shape of the driving field $E(t)$}

\begin{figure}[htb]
\begin{center}
%\centerline{\epsfig{file=fig3.eps,width=2.6in,height=2.0in,angle=0}}
\leavevmode
\epsfxsize=3.in
\epsfbox{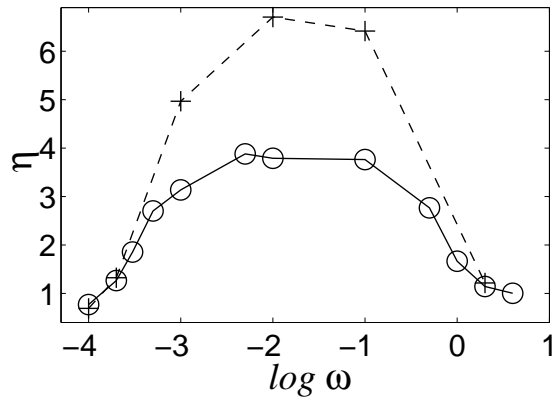}
\end{center}
\caption{Dependence of the amplification of the jump rate $\eta$ 
on the logarithm of the
driving frequency for $\gamma=1$, $D=0.03$ and $A=0.09$ for different 
shapes of the driving field $E(t)$: $\mu=0$ ($\circ$) and $\mu=10$ ($+$).}
\label{fig3a}
\end{figure}

Finally we discuss the generality of the obtained results with respect
to the shape of the driving force $E(t)$ since the 
distribution of the residence times $t_{\pm}$ on the initial phase has
been omitted when the adiabatic limit has been considered.  To check this,
numerical simulations have been performed for the two cases of the
driving field $E(t)$. In the Fig. \ref{fig3a} we show the dependence of the
amplification ratio when the drive is sinusoidal ($\mu=0$) and when
it is almost stepwise ($\mu=10$). In the latter case one can assume the
distribution of the residence times along the halfperiod of the oscillation
of the driving force to be independent on the phase $\phi$. 
No qualitative differences can be observed
for these cases, only the quantitative ones. In the
adiabatic limit, the amplification factor is less then one and 
increases with the growth of the
driving frequency regardless of the shape of the driving field $E(t)$. 

Another interesting observation is that the maximal amplification of the
jump rate depends on the shape of the driving field. We see that for
the stepwise shape (large $\mu$) we obtain much better amplification.
If $E(t)$ is of the stepwise shape, the time intervals when the 
potential wells are almost not desymmetrized, are very short. In other words,
almost all the time one well is in the uppermost position while another
one is in the lowest. The time intervals when the potential
is almost symmetric do not contribute anything into the observed resonance,
since in that case the jump processes occur with the rate close to $r_k$.
Thus minimization of these time intervals effectively must enhance
the jumps and
%@@@@@@@@@@@@@@@@@@@@@@@@@@@@@@@@@@@@@@@@@@@@@@@@@@@@@@@@@@@@@@@@@@@@@@
%
%-------------------Figure 6-------------------------------------------
%@@@@@@@@@@@@@@@@@@@@@@@@@@@@@@@@@@@@@@@@@@@@@@@@@@@@@@@@@@@@@@@@@@@@@@
 selecting appropriately the shape of $E(t)$ one might increase
the amplification significantly. However, this topic is beyond 
the scope of our paper.

\section{Conclusions}

To summarize our results, we have found that the jump rate of an
overdamped Brownian particle in the double-well potential, driven 
by a periodic force with zero mean depends resonantly on the 
frequency of the drive. This phenomenon exists for any ratio of 
the driving amplitude $A$ to the noise strength, $D$,
however is very well pronounced if $A > D$. We also have observed 
that this phenomenon occurs for different shapes of the 
driving force $E(t)$. For the stepwise function $E(t)$ the jump rate
can be twice larger then in the case of the sinusoidal drive with 
the same amplitude. An interesting result has been observed in
Ref. \cite{gdh96prl}, where diffusion of
an overdamped particle in the spatially periodic potential
has been studied. The particle is also under the influence of the
time-periodic force. Among other results, authors have noticed
``acceleration'' of diffusion for certain values of the 
driving period. However, while it was only in the limit
$A/D \gg 1$ they observe the effect, in our case
it happens for any ratio of $A/D$.

We also would like to mention some analogy of the observed
phenomenon with the resonant activation over a fluctuating barrier,
which can be driven either randomly as in Ref. \cite{dg92prl} or
deterministically via periodic force (see Ref. \cite{ps00pla}).
These papers have shown that the mean first-passage time (MFPT) as
a function of the intensity (rate, frequency) of the barrier fluctuations 
displays a local minimum. In our problem the driving force $E(t)$  
effectively induces the oscillating change of the depth of the wells 
and the average 
number of jumps $\bar n$ is nothing but an inverse MFPT.

Finally we would like to stress the importance of the {\it overdamped}
case since
the dynamics of many biological objects is modeled by
overdamped systems, so the above problem could have a wide 
range of applications.

This work has been supported by 
the European Union grant LOCNET project no. HPRN-CT-1999-00163.
Y. Z. also wishes to acknowledge partial financial support from the
Ukrainian Fundamental Research Fund, contract no. 0103U007750.

%----------------------------------------------------------------------
% References
%----------------------------------------------------------------------   

\end{document}